\documentclass[prd,aps,showpacs,nofootinbib,preprint,tightenlines]{revtex4}

\newcommand{\checked}[1]{}
\newcommand{\comment}[1]{}

\usepackage{graphicx}
\usepackage{rotating}
\usepackage{axodraw}
\newcommand{\beq}{\begin{equation}}
\newcommand{\eeq}{\end{equation}}
\newcommand{\bqa}{\begin{eqnarray}}
\newcommand{\eqa}{\end{eqnarray}}

\newcommand{\picb}[1]{\;\parbox[c]{48pt}{\begin{picture}(45,30)(-9,0)
\SetWidth{1.0}\SetScale{1.0} #1 \end{picture}}\;}

\def\Lwidth{1}

\def\Agl(#1,#2)(#3,#4,#5){\PhotonArc(#1,#2)(#3,#4,#5){\Lwidth}
{6.283 #3 mul 360 div #4 #5 sub #4 #5 sub mul sqrt mul Ldensity mul}}
\def\Lgl(#1,#2)(#3,#4){\Photon(#1,#2)(#3,#4){\Lwidth}
{#1 #3 sub #1 #3 sub mul #2 #4 sub #2 #4 sub mul add sqrt Ldensity mul}}
\def\Agh(#1,#2)(#3,#4,#5){\DashArrowArc(#1,#2)(#3,#4,#5){1}}
\def\Aagh(#1,#2)(#3,#4,#5){\DashArrowArcn(#1,#2)(#3,#5,#4){1}}
\def\Lgh(#1,#2)(#3,#4){\DashArrowLine(#1,#2)(#3,#4){1}}
\def\Lagh(#1,#2)(#3,#4){\DashArrowLine(#3,#4)(#1,#2){1}}
\def\Ahh(#1,#2)(#3,#4,#5){\DashCArc(#1,#2)(#3,#4,#5){1}}
\def\Lhh(#1,#2)(#3,#4){\DashLine(#1,#2)(#3,#4){1}}
\def\Aqu(#1,#2)(#3,#4,#5){\ArrowArc(#1,#2)(#3,#4,#5)}
\def\Aaqu(#1,#2)(#3,#4,#5){\ArrowArcn(#1,#2)(#3,#5,#4)}
\def\Lqu(#1,#2)(#3,#4){\ArrowLine(#1,#2)(#3,#4)}
\def\Laqu(#1,#2)(#3,#4){\ArrowLine(#3,#4)(#1,#2)}
\def\Aqq(#1,#2)(#3,#4,#5){\CArc(#1,#2)(#3,#4,#5)}
\def\Lqq(#1,#2)(#3,#4){\ArrowLine(#1,#2)(#3,#4)}
\def\Asc(#1,#2)(#3,#4,#5){\ArrowArc(#1,#2)(#3,#4,#5)}
\def\Lsc(#1,#2)(#3,#4){\ArrowLine(#1,#2)(#3,#4)}
\def\DAsc(#1,#2)(#3,#4,#5){\DashCArc(#1,#2)(#3,#4,#5){3}}
\def\DLsc(#1,#2)(#3,#4){\DashLine(#1,#2)(#3,#4){3}}
\def\TAsc(#1,#2)(#3,#4,#5){\SetWidth{2.0}\CArc(#1,#2)(#3,#4,#5)\SetWidth{1.0}}
\def\TLsc(#1,#2)(#3,#4){\SetWidth{2.0}\ArrowLine(#1,#2)(#3,#4)\SetWidth{1.0}}

\begin{document}

\title{Collisional Energy Loss of a Heavy Quark in an Anisotropic Quark-Gluon Plasma}

\preprint{ TUW-04-19 }

\author{Paul Romatschke}
\author{Michael Strickland}
\affiliation{Institut f\"ur Theoretische Physik, Technische Universit\"at Wien,
	Wiedner Hauptstrasse 8-10, A-1040 Vienna, Austria
     \vspace{1cm}
	}

\begin{abstract}

We compute the leading-order collisional energy loss of a heavy quark 
propagating through a quark-gluon plasma in which the quark and gluon 
distributions are anisotropic in momentum space. Following the calculation 
outlined for QED in an earlier work we indicate the differences encountered in 
QCD and their effect on the collisional energy loss results.  For a 20 GeV 
bottom quark we show that momentum space anisotropies can result in the 
collisional heavy quark energy loss varying with the angle of propagation by up 
to 50\%.  For low velocity quarks we show that anisotropies result in energy 
gain instead of energy loss with the energy gain focused in such a way as to 
accelerate particles along the anisotropy direction thereby reducing the 
momentum-space anisotropy.  The origin of this negative energy loss is 
explicitly identified as being related to the presence of plasma instabilities 
in the system.

\end{abstract}
\maketitle
\newpage

\small

\section{Introduction}

An understanding of the production, propagation, and hadronization of heavy 
quarks in relativistic heavy-ion collisions is important for predicting a number 
of experimental observables including the heavy-meson spectrum, the single 
lepton spectrum, and the dilepton spectrum.  The first experimental results for 
the inclusive electron spectrum have been reported \cite{adcox:2002} in addition 
to the first measurements of $J/\psi$ production at RHIC \cite{nagle:2002}.  The 
measurement of the inclusive electron spectrum allows for a determination of 
heavy quark energy loss since it is primarily due to the semi-leptonic decay of 
charm quarks.  The heavy quark energy loss comes into play since it is 
necessary in order to predict the heavy quark energy at the decay point. It is 
therefore important to have a thorough theoretical understanding of heavy 
quark energy loss for a proper comparison with the experimental results.  In 
this paper we will show that in QCD there is a modification of the leading-order 
(collisional) heavy quark energy loss if there is a momentum-space anisotropy 
in the underlying quark and gluon distribution functions. 

Here we will focus on the collisional energy loss using the techniques of 
Braaten and Thoma \cite{Braaten:1991jj,Braaten:1991we} which have been extended 
to anisotropic systems in Ref.~\cite{Romatschke:2003vc}.  For a review of the 
radiative energy loss in isotropic systems see Refs.~\cite{Baier:2000mf} and 
\cite{Accardi:2003gp}.  The method employed by Braaten and Thoma gives the complete 
result for the collisional energy loss by considering independently the 
contributions from soft (involving momenta $q\sim m_D$) and hard ($q\sim T$) 
gluon exchange. The two scales, $m_D$ and $T$, are then separated by an 
arbitrary momentum scale $q^{*}$ which cuts off the UV and IR divergences in the 
soft and hard contributions, respectively. Moreover, it was found that in the 
weak-coupling or high-temperature limit $\alpha_s\ll 1$ the condition $m_D\ll 
q^* \ll T$ could be used to expand the resulting integral expressions for the 
soft and hard contributions further, giving an analytic result for the energy 
loss independent of the separation scale 
$q^*$~\cite{Braaten:1991jj,Braaten:1991we}. However, in the case of QCD at 
temperatures accessible by heavy-ion collision experiments the coupling constant 
is quite large ($\alpha_s \sim 0.2-0.3$) and as a consequence the analytic 
expression of Braaten and Thoma can give unphysical results for the 
energy loss as we will demonstrate.

On the good news side, it turns out that when one does not expand the integral 
expressions with respect to the condition $m_D\ll q^* \ll T$ the isotropic 
contributions to the energy loss \emph{always} stay positive, even for very 
large coupling, as has been shown for QED \cite{Romatschke:2003vc}. However, 
this comes at the expense of giving up independence of the complete result on 
the scale $q^*$, which then has to be fixed somehow. Fortunately, it turns out 
that in the weak-coupling limit this scale dependence becomes very small and as 
a consequence, unless $q^*$ is taken to be very large ($q^*\gg T$) or very small 
($q^*\ll m_D$), one recovers the analytic results found by Braaten and Thoma. 
When the coupling is increased one can then still fix $q^*$ by the ``principle 
of minimal sensitivity'', which means that $q^*$ is chosen such that the energy 
loss (and therefore also the variation with respect to $q^*$) is minimized. 
In addition, 
the principle of minimal sensitivity provides a way to estimate the 
theoretical uncertainty of the result by varying $q^*$ by a fixed amount around 
the point at which the energy loss is least sensitive to this scale.

The main reason why a calculation of this type is important to perform is that 
the presence of plasma instabilities \cite{Weibel:1959,Mrowczynski:1993qm,%
Mrowczynski:1994xv,Mrowczynski:1997vh,Randrup:2003cw,Romatschke:2003ms,%
Romatschke:2003vc,Arnold:2003rq,Mrowczynski:2004mr,Romatschke:2003yc,%
Romatschke:2004jh,Arnold:2004al} in anisotropic systems could have a significant 
effect on observables like the heavy quark energy loss.  In fact, the presence 
of such instabilities naively renders the calculation of the soft part 
divergent; however, there is protection mechanism dubbed ``dynamical shielding'' 
which renders the collisional energy loss finite for QED 
\cite{Romatschke:2003vc}. This protection mechanism trivially extends to the 
case of QCD since the only thing that changes in going from soft QED to soft QCD 
at leading-order is the numerical value of the Debye mass. However, at large 
coupling the presence of instabilities and associated poles on unphysical 
Riemann sheets \cite{Romatschke:2004jh} causes significant changes in the soft 
energy loss contribution for both QED and QCD.  In fact, as we will discuss, the 
unphysical poles can even change the sign of the heavy quark energy loss at low 
momentum turning it instead into energy \emph{gain}.

The paper is organized as follows: In Sec.~\ref{setuprep} we review the 
calculation of the collisional energy loss including the case of very strong 
anisotropies. In Sec.~\ref{sec:umodes} we discuss the role that modes on the 
unphysical sheets play in the calculation. In Sec.~\ref{QCDEloss} we present 
results for the collisional energy loss of a heavy quark for isotropic and 
anisotropic systems.  We give our conclusions in Sec.~\ref{Conc}.

\section{Collisional energy loss in QCD}
\label{setuprep}

Following Refs.~\cite{Braaten:1991jj,Braaten:1991we,%
Romatschke:2003vc,Romatschke:2003yc} the collisional energy loss of a heavy 
quark with velocity $\textbf{v}$ is divided into a soft contribution and a hard 
contribution.  The soft contribution within QCD has exactly the same form as in 
QED except that the expression for the Debye mass is different and there is an
overall Casimir which has to be taken into account.  Therefore the soft 
contribution from Ref.~\cite{Romatschke:2003vc} can be taken directly and only 
needs to be rescaled appropriately.  The hard contribution in QCD is, however, 
different for two reasons:  a new gluon scattering diagram appears due to the 
three-gluon coupling and the graphs which are topologically equivalent to the 
QED fermion-photon graphs no longer cancel due to non-trivial Casimir 
invariants. In the next two subsections we present explicit integral expressions 
for both the soft and hard contributions to the energy loss in the limit of 
infinite quark mass.  In the final subsection we discuss how the limit of 
infinite anisotropy can be taken.

\subsection{Soft contribution}

The soft contribution in the case of QCD is given by the expression 
\beq
-\left({{\rm d} W\over {\rm d} t}\right)_{\rm soft} = {\mathcal Q}^2 %
  \; {\rm Im} \int\!{d^3 {\bf q} \over (2 \pi)^3} \, ({\bf q}\cdot{\bf v}) \, v^i %
\left[\Delta^{ij}(Q)-\Delta^{ij}_0(Q)\right] v^j  \; ,
\label{eloss2}
\eeq
\checked{m}
where ${\mathcal Q}$ is the quark color charge in the fundamental representation
with ${\mathcal Q}^2= g^2 (N_c^2-1)/(2 N_c)$ and the two propagators 
$\Delta^{ij}(Q)$ and $\Delta^{ij}_0(Q)$ denote the hard-loop  
and free gluon propagators, respectively.  The  hard-loop
gluon propagator in an anisotropic system has been calculated in
Ref.~\cite{Romatschke:2003ms} and can be expressed as
\beq
{\bf \Delta}(Q) =  \Delta_A(Q) \, [{\bf A}-{\bf C}] 
        + \Delta_G(Q) \, [(q^2 - \omega^2 + \alpha + \gamma) {\bf B} + 
        (\beta-\omega^2) {\bf C}  - \delta {\bf D}] \; ,
\eeq
\checked{mp}
where the tensor basis for the spacelike components of a system with one preferred
direction specified by $\hat{\bf n}$ is: $A^{ij}=\delta^{ij}-k^{i}k^{j}/k^2$, $B^{ij}=k^{i}k^{j}/k^2$,
$C^{ij}=\tilde{n}^{i} \tilde{n}^{j} / \tilde{n}^2$, $D^{ij}=k^{i}\tilde{n}^{j}+k^{j}\tilde{n}^{i}$
with $\tilde n^{i}=A^{ij} n^{j}$.  The propagators $\Delta_A$ and $\Delta_G$ are then given by
\bqa
\Delta_A^{-1}(Q) &=& q^2 - \omega^2 + \alpha \; , \nonumber \\
\Delta_G^{-1}(Q) &=& (q^2 - \omega^2 + \alpha + \gamma)(\beta-\omega^2)-q^2 \tilde n^2 \delta^2 \; ,
\label{propfnc}
\eqa
\checked{mp}
with $\alpha$, $\beta$, $\gamma$, and $\delta$ being the hard-loop gluon 
structure functions.  Except for some special cases (certain angles of propagation, 
infinite anisotropy, etc.), the analytic expressions for the structure functions are 
quite complicated in form and it is more convenient to use a numerical 
evaluation of their integral representations \cite{Romatschke:2003ms}. 

The structure functions of anisotropic
systems depend on the precise form of the quark and gluon distribution functions.
Here we will assume that the distribution function is spatially homogeneous
and therefore given by
\bqa
f({\bf p}) \equiv 2 N_f \left(n({\bf p}) + \bar n ({\bf p})\right) + 4 N_c n_g({\bf p}) \; ,
\label{distfncs}
\eqa
\checked{mp}
where $n$, $\bar n$, and $n_g$ are the distribution functions of quarks, anti-quarks, 
and gluons, respectively.\footnote{There is a factor two difference 
between Eq.~(\ref{distfncs}) and the equivalent expression given in 
Ref.~\cite{Romatschke:2003vc}. We have changed our notation in this work so that 
all symmetry factors are specified in $f({\bf p})$.  Final results are 
unaffected by this redefinition.} We will further assume that the anisotropic 
distribution function, $f({\bf p})$, can be obtained from an isotropic 
distribution function by rescaling one direction in momentum space

\beq
f({\bf p}) =  N(\xi)\,f_{\rm iso}\left(p \sqrt{1+\xi(\hat{\bf p} \cdot \hat{\bf n})^2}\right) \, ,
\label{distfunc}
\eeq
\checked{mp}
where $N(\xi)=\sqrt{1+\xi}$ is a normalization constant with $\xi$ the strength 
of the anisotropy.  The anisotropy strength $\xi$ is in the range $-
1<\xi<\infty$ with $\xi=0$ corresponding to the original isotropic distribution. 
In order to simplify the analysis we take quarks and gluons to have the same 
anisotropy parameter although in general they can be different. Even though its 
effects might be interesting, different anisotropies for quarks and gluons will 
not be studied in this paper. 

The soft contribution to the collisional energy loss
of a heavy quark in an anisotropic quark-gluon plasma is then given by
\bqa
-\left({{\rm d} W\over {\rm d} x}\right)_{\rm soft}\!\!\!\!\!\!\! &=& %
\!\!\! \frac{{\mathcal Q}%
^2}{v}  \, {\rm Im} \, %
\int {d \Omega_{q}\over (2 \pi)^3} \, %
\frac{\hat{\omega}}{(1-\hat{\omega}^2)} \, %
\Biggl[-\alpha %
\frac{(v^2-\hat{\omega}^2-{(\tilde{\bf n}\cdot{\bf v})^2\over\tilde{n}^2})}
{2(1-\hat{\omega}^2)}%
\ln{\frac{q^{* 2}(1-\hat{\omega}^2)+\alpha}{\alpha}} \nonumber \\
&& \hspace{5.5cm} + F(q^{\star})-F(0)\Biggr] \,,
\label{Elosssoftfinal}
\eqa
\checked{mp}
where
\beq
F(q)=\frac{\mathcal{A}}{4 \mathcal{C}} \ln{\left(%
-4\mathcal{C}\left(\mathcal{C} q^4+\mathcal{D}q^2+\mathcal{E}\right)%
\right)}
+\frac{\mathcal{AD}-2\mathcal{BC}}{4 %
\mathcal{C}\sqrt{\mathcal{D}^2-4 \mathcal{CE}}}\ln{\frac{%
\sqrt{\mathcal{D}^2-4 \mathcal{CE}}+\mathcal{D}+2\mathcal{C}q^2}{
\sqrt{\mathcal{D}^2-4 \mathcal{CE}}-\mathcal{D}-2\mathcal{C}q^2}} \; ,
\eeq
\checked{mp}
with
\bqa
\mathcal{A}&=&(1-\hat{\omega}^2)^2 \beta+\hat{\omega}^2%
{(\tilde{\bf n}\cdot{\bf v})^2\over\tilde{n}^2} (\alpha+\gamma)-2 \hat{%
\omega}(1-\hat{\omega}^2)(\tilde{\bf n}\cdot{\bf v}) \hat{\delta} \nonumber \; ,\\
\mathcal{B}&=&\left((\alpha+\gamma)\beta-\tilde{\bf{n}}^2\hat{\delta}^2\right)%
(1-\hat{\omega}^2-{(\tilde{\bf n}\cdot{\bf v})^2\over\tilde{n}^2})\nonumber \; ,\\
\mathcal{C}&=&-\hat{\omega}^2(1-\hat{\omega}^2)\nonumber \; , \\
\mathcal{D}&=&-\hat{\omega}^2(\alpha+\gamma)+(1-\hat{\omega}^2)\beta \; , \nonumber \\
\mathcal{E}&=&(\alpha+\gamma)\beta-\tilde{\bf{n}}^2\hat{\delta}^2 \; , 
\label{aeeqs}
\eqa
\checked{mp}
and $\hat{\omega} = {\bf \hat{q}}\cdot{\bf v}$, and $\hat{\delta}= q \delta$.
In order to regulate the soft contribution it is necessary to introduce a 
UV momentum cutoff $q^{*}$ on the $q$ integration. 

Note that in the case of an anisotropic system unstable modes are present that 
manifest themselves as potentially unregulated poles of the propagator in the 
static limit. However, as discussed in our previous paper 
\cite{Romatschke:2003vc} the mechanism of ``dynamical shielding'' protects the 
collisional energy loss from these potentially devastating singularities in the 
soft contribution.  These singularities could arise in terms containing, for 
example, $(q^2 + \alpha)^{-1}$, due to the fact that in the static limit the 
structure function $\alpha$ is negative-valued.  However, if one takes the static 
limit of $\alpha$ carefully we find
\bqa
\lim_{\omega\rightarrow0} \alpha(\omega,q) = M^2(-1 + i D \hat\omega) 
	+ {\cal O}(\omega^2) \; ,
\eqa
where $M$ and $D$ depend on the angle of propagation with respect to the 
anisotropy vector and the strength of the anisotropy.  As long as $D$ is non-vanishing 
the singularities are regularized because of the $\hat\omega$ in the 
numerator of Eq.~(\ref{eloss2}) and we call the singularity 
``dynamically shielded''. A proof of dynamical shielding for very weak and 
strong anisotropies can be found in Refs.~\cite{Romatschke:2003vc} and 
\cite{Romatschke:2003yc}.

In the isotropic limit ($\xi=0$) Eq.~(\ref{Elosssoftfinal}) corresponds 
to the result of Thoma and Gyulassy \cite{Thoma:1991fm} and one obtains the soft 
contribution found by Braaten and Thoma \cite{Braaten:1991we} when expanding 
Eq.~(\ref{Elosssoftfinal}) under the additional assumption $q^{*}\gg m_D$ 
\cite{Romatschke:2003yc}. The effect of this expansion is that the result of 
Ref.~\cite{Braaten:1991we} becomes negative for small $q^{*}/m_D$ (but allows 
the calculation of an analytic result for the isotropic collisional energy loss) 
while the unexpanded result is positive for all $q^{*}/m_D$.

\subsection{Hard Contribution}

\begin{figure}[t]
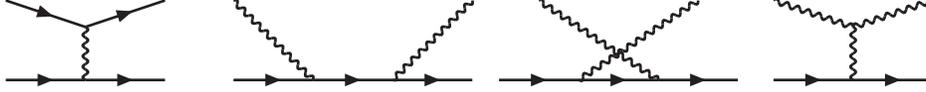

\bqa \nonumber
\picb{\Lqu(0,0)(30,0) \Lqu(30,0)(60,0) \Lgl(30,20)(30,0) \Lqu(0,30)(30,20)%
 \Lqu(30,20)(60,30)}~\hspace{1cm}
\picb{\Lqu(0,0)(30,0) \Lqu(30,0)(60,0) \Lqu(60,0)(90,0) \Lgl(0,30)(30,0)%
 \Lgl(90,30)(60,0)}~
\hspace{1.5cm}
\picb{\Lqu(0,0)(30,0) \Lqu(30,0)(60,0) \Lqu(60,0)(90,0) \Lgl(75,30)(30,0)%
 \Lgl(15,30)(60,0)}
\hspace{1.75cm}
\picb{\Lqu(0,0)(30,0) \Lqu(30,0)(60,0) \Lgl(30,20)(30,0) \Lgl(0,30)(30,20)%
 \Lgl(30,20)(60,30)}
~\hspace{1cm}
\eqa
\caption[a]{%
Tree-level Feynman diagrams for the scattering processes 
$Q q\rightarrow Q q$ (first diagram) and 
$Q g \rightarrow Q g$ (remaining diagrams).}
\label{hard-diagrams}
\end{figure}

The hard contribution can be separated into two parts: one contribution
coming from the scattering of the heavy quark on quarks in the plasma
and another that takes into account the scattering on plasma gluons (corresponding
to the tree-level diagrams shown in Fig.~\ref{hard-diagrams}).  Assuming
the velocity of the quark to be much higher then the ratio of the plasma
temperature to the energy of the quark, $v\gg T/E$, the contribution
coming from quark-quark scattering can be reduced to \cite{Braaten:1991we}
\bqa
-\left({\rm d} W \over {\rm d} x\right)_{\rm hard}^{Qq} \!\! &=&%
 {2 (4 \pi)^3 N_f \alpha_s^2 \over 3v}
        \int {d^3{\bf k} \over (2 \pi)^3} {n({\bf k})\over k} 
        \int {d^3{\bf k^\prime} \over (2 \pi)^3} {1 - n({\bf k}^\prime) %
\over k^\prime}
        \delta(\omega-{\bf v}\cdot{\bf q}) \nonumber \\
&&       \hspace*{-7mm}\times \; \Theta(q-q^*) {\omega \over (\omega^2-q^2)^2} 
        \left[ 2 (k-{\bf v}\cdot{\bf k})^2%
                + {1 - v^2 \over 2} (\omega^2 - q^2) \right], \;
\label{hQq}
\eqa
\checked{mp}
after performing the Dirac traces and evaluating the sum over spins.
The contribution coming from quark-gluon scattering gives
\bqa
-\left({\rm d} W \over {\rm d} x\right)_{\rm hard}^{Qg} \!\!&=&%
 {(4 \pi)^3 \alpha_s^2 \over 2v}
        \int {d^3{\bf k} \over (2 \pi)^3} {n_{g}({\bf k})\over k} 
        \int {d^3{\bf k^\prime} \over (2 \pi)^3} {1 + n_{g}({\bf k}^\prime) %
\over k^\prime}
        \delta(\omega-{\bf v}\cdot{\bf q}) \Theta(q-q^*) \nonumber \\
 &&       \times \; \omega 
        \left[ \frac{(1-v^2)^2}{(k-{\bf v}\cdot{\bf k})^2}+ 8 \frac{%
(k-{\bf v}\cdot{\bf k})^2+\frac{1-v^2}{2}(\omega^2-q^2)}{(\omega^2-q^2)^2}
\right].\quad
\eqa
\checked{mp}
Here $n({\bf k})$ and $n_{g}({\bf k})$ 
are the anisotropic versions of the tree-level Fermi-Dirac and 
Bose-Einstein distribution functions at zero chemical potential specified
in Eq.~(\ref{distfncs}), 
$\omega = k^\prime - k$, and ${\bf q} = {\bf k}^\prime - {\bf k}$. Note
also that $q^*$ acts as an IR cutoff for the $q$ integration. Since 
the integrand is odd under the interchange 
${\bf k} \leftrightarrow {\bf k}^\prime$ the terms involving the products 
$n({\bf k})n({\bf k}^\prime)$ and 
$n_{g}({\bf k})n_{g}({\bf k}^\prime)$ vanish
since they are symmetric. Following Ref.~\cite{Romatschke:2003vc}, some
of the integrations can be performed giving
%
\bqa
-\left({\rm d} W \over {\rm d} x\right)_{\rm hard}^{Qq} &\!\!\!=\!\!& %
{8 \alpha_s^2 N_f (\hat{q}^{*})^2 T^2 \sqrt{1+\xi} \over 3 \pi^3 v}
        \int_{0}^{\infty} z \, dz \int_{-1}^{1} d\cos \theta_k 
        \left[\int_{0}^{2 \pi} \! d \phi_k F_1 (x)\right] %
       \nonumber \\
        && \hspace{-1cm} \times \int_{-1}^{1} d\cos{\theta_{q}} %
     		\; {\mathcal T}
        {v \cos \theta_q \over (v^2 \cos^2 \theta_q-1)^2} 
        \left[ 2 z^2(1-v \cos \theta_k)^2  + {1 - v^2 \over 2} (v^2 \cos^2 \theta_q - 1) \right], %
\nonumber\\
-\left({\rm d} W \over {\rm d} x\right)_{\rm hard}^{Qg} &\!\!\!=\!\!& %
{2 \alpha_s^2 (\hat{q}^{*})^2 T^2 \sqrt{1+\xi} \over \pi^3 v}
        \int_{0}^{\infty} z \, dz \int_{-1}^{1} d\cos \theta_k 
        \left[\int_{0}^{2 \pi} \! d \phi_k F_2 (x)\right] %
       \nonumber \\
        && \hspace{-1cm} \times \int_{-1}^{1} d\cos{\theta_{q}} %
      \; {\mathcal T} v \cos \theta_q
 \left[\frac{(1-v^2)^2}{z^2(1-v \cos \theta_k)^2}
+8 \frac{z^2(1-v \cos \theta_k)^2+\frac{1-v^2}{2}(v^2 \cos^2 \theta_q-1)}
{(v^2 \cos^2 \theta_q-1)^2}\right],\nonumber\\
\label{myelosshard3}
\eqa
\checked{mp}
where ${\mathcal T}$ denotes the unwieldy expression
\beq
{\mathcal T}=        \Theta(z+v \cos{\theta_q})
        \frac{\Theta(4 z^2 \sin^2 \theta_k \sin^2 \theta_q-(1-v^2 \cos^2 \theta_
q+%
        2 \cos \theta_q z (\cos \theta_k- v))^2)}%
        {\sqrt{4 z^2 \sin^2 \theta_k \sin^2 \theta_q-(1-v^2 \cos^2 \theta_q+%
        2 \cos \theta_q z (\cos \theta_k- v))^2}},
\eeq
\checked{mp}
and
\bqa
F_1(x) &=& \frac{1}{x^2 T^2}\int_{xT}^{\infty} dq\, q \ n(q) = %
\frac{x \ln(1+\exp{(-x)})-{\rm Li}_2(-\exp{(-x)})}{x^2}
\label{F11} \nonumber \\
F_2(x) &=& \frac{1}{x^2 T^2}\int_{xT}^{\infty} dq\, q \ n_g(q) = %
\frac{-x \ln(1-\exp{(-x)})+{\rm Li}_2(\exp{(-x)})}{x^2}  \, , 
\eqa
\checked{mp}
where 
$x=q^{*} z \sqrt{1+\xi(n_x \sin\theta_k \cos\phi_k+n_z%
 \cos\theta_k)^2}/T$,
with  $n_z=\bf{\hat{n}}\cdot\hat{\bf v}$ and \hbox{$1=n_x^2+n_z^2$}.
The remaining integrations have to be performed numerically.

Similar to what has been found for the soft part, in the isotropic
limit the hard contribution becomes that of Braaten and Thoma 
\cite{Braaten:1991we} when expanding Eq.~(\ref{myelosshard3}) with respect to
$q^{*}\ll T$; also, the expanded result becomes negative for large
$q^{*}/T$, while the unexpanded result is always positive in the isotropic
limit.

\subsection{Very strong anisotropies}

There are however
a few special cases where the structure functions become quite simple, notably
the extreme anisotropy limit, $\xi\rightarrow\infty$, which corresponds to
the parton distribution function having the form
\beq
 f({\bf p}) = F(p_\perp) \delta(p_z) \, .
\label{deltalimit}
\eeq
\checked{mp}
The resulting expressions for the structure functions have been calculated in 
Ref.~\cite{Romatschke:2003yc} and this special case has been treated in more 
detail in a separate publication \cite{Romatschke:2004jh}. Using these structure 
functions one is able to calculate the soft contribution to the energy loss in 
the large $\xi$ case by evaluation of Eq.~(\ref{Elosssoftfinal}). However, in 
the extreme limit $\xi\rightarrow \infty$ this is complicated by the presence of 
troublesome spacelike quasiparticle modes \cite{Romatschke:2004jh}.  For large but
finite $\xi$ these modes are integrable because they move off the physical sheet 
onto the neighboring unphysical sheets but for $\xi=\infty$ they result in a 
pole in the integrand that needs to be regulated. Fortunately, a natural 
regularization is provided by the fact that for any finite $\xi$, these modes 
are outside of the physical spectrum; therefore, when keeping $\xi$ finite and 
sending it to infinity only after having done the integrations, one is able to 
produce well-defined results.\footnote{In practice it is 
more convenient to use the $\xi\rightarrow \infty$ expressions for the structure 
functions and regulate the integrand ``by hand'' so that it agrees with the 
finite $\xi$ case.}

The hard contribution in the large $\xi$ case is easily found by using
Eq.~(\ref{myelosshard3}) together with
\beq
\lim_{\xi\rightarrow \infty} \sqrt{1+\xi} \, F_1\left(\hat{q}^{*} z \sqrt{1+ \xi%
(\hat{k}\cdot \hat{n})^2}\right)=\delta(\hat{k}\cdot \hat{n})
\int_{-\infty}^{\infty} dy \, F_1(\hat{q}^{*} z \sqrt{1+y^2})
\eeq
\checked{mp}
and similarly for $F_2$. Using 
\beq
\hat{k}\cdot \hat{n}=n_x\sin{\theta_k} \cos{\phi_k}+ n_z\cos{\theta_k} \, ,
\eeq
\checked{mp}
the delta function can be cast into the form
\beq
\delta(\hat{k}\cdot \hat{n})=\frac{\delta(\phi_k-\phi_0)}{\sqrt{n_x^2 %
\sin^2{\theta_k}-n_z^2 \cos^2{\theta_k}}} \, ,
\eeq
\checked{mp}
with 
\beq
n_x \cos \phi_0 = - n_z {\rm cot} \theta_k \, .
\eeq
Using the symmetry $2\pi-\phi\leftrightarrow \phi$ and substituting
$\cos{\theta_k}\rightarrow n_x \cos{\theta_k}$ the remaining integrations
are easily implemented.

\section{Effects of modes on the unphysical sheet}
\label{sec:umodes}

In a previous paper we have discussed the presence of nearly-spacelike 
``unphysical modes'' in anisotropic plasmas~\cite{Romatschke:2004jh}. In that 
work we showed that for large anisotropies new relevant singularities are 
present on neighboring unphysical Riemann sheets.  The presence of these 
unphysical poles has a resonance-like effect on propagators on the physical 
sheet and therefore can influence observables which are sensitive to spacelike 
momentum. Before presenting our numerical results we would like to first discuss 
how the presence of these unphysical modes affects the soft contribution to the 
energy loss. In order to do this as clearly as possible we consider the case 
when the heavy quark is propagating along the anisotropy direction, i.e. 
$\theta_v=0$, in which case the azimuthal integration in Eq.~(\ref{eloss2}) 
becomes trivial since $\hat\omega \rightarrow v \cos{\theta_q}$.  In this case 
the soft contribution to the energy loss takes the form
\bqa
-\left({{\rm d} W\over {\rm d} x} \right)_{\rm soft,\theta_v=0} &=&%
\frac{{\mathcal Q}%
^2}{(2 \pi)^2 v}
\, {\rm Im} \, \int_{0}^{\pi} d\theta_q \,%
\int_{0}^{q^*} dq \; \frac{q \, \hat\omega \sin{\theta_q}}{1-\hat\omega^2} \, 
\frac{q^2 \mathcal{A}+\mathcal{B}}{q^4 \mathcal{C}+q^2 \mathcal{D}+ \mathcal{E}}
\, ,
\label{elossReg2}
\eqa
\checked{mp}
with $\tilde{n}^2 = \tilde{\bf n}\cdot\hat{\bf v} = 1 - \cos^2\theta_q$ 
understood in Eq.~(\ref{aeeqs}).
In this special case, we can easily relate Eq.~(\ref{elossReg2}) to
the collective modes in the system: this is done by investigating the integrand
at fixed angles $\theta_q$, which represents the contribution from the straight
line $\omega=q\,v \cos{\theta_q}$ from $q=0$ to $q=q^*$. 
This integration path can then be directly compared to the unphysical mode
dispersion relation $\omega_i=\omega_i(q,\theta_q)$ where $i=\{A,B\}$ indexes
the unphysical mode in question.

\begin{figure}

\includegraphics[width=14.8cm]{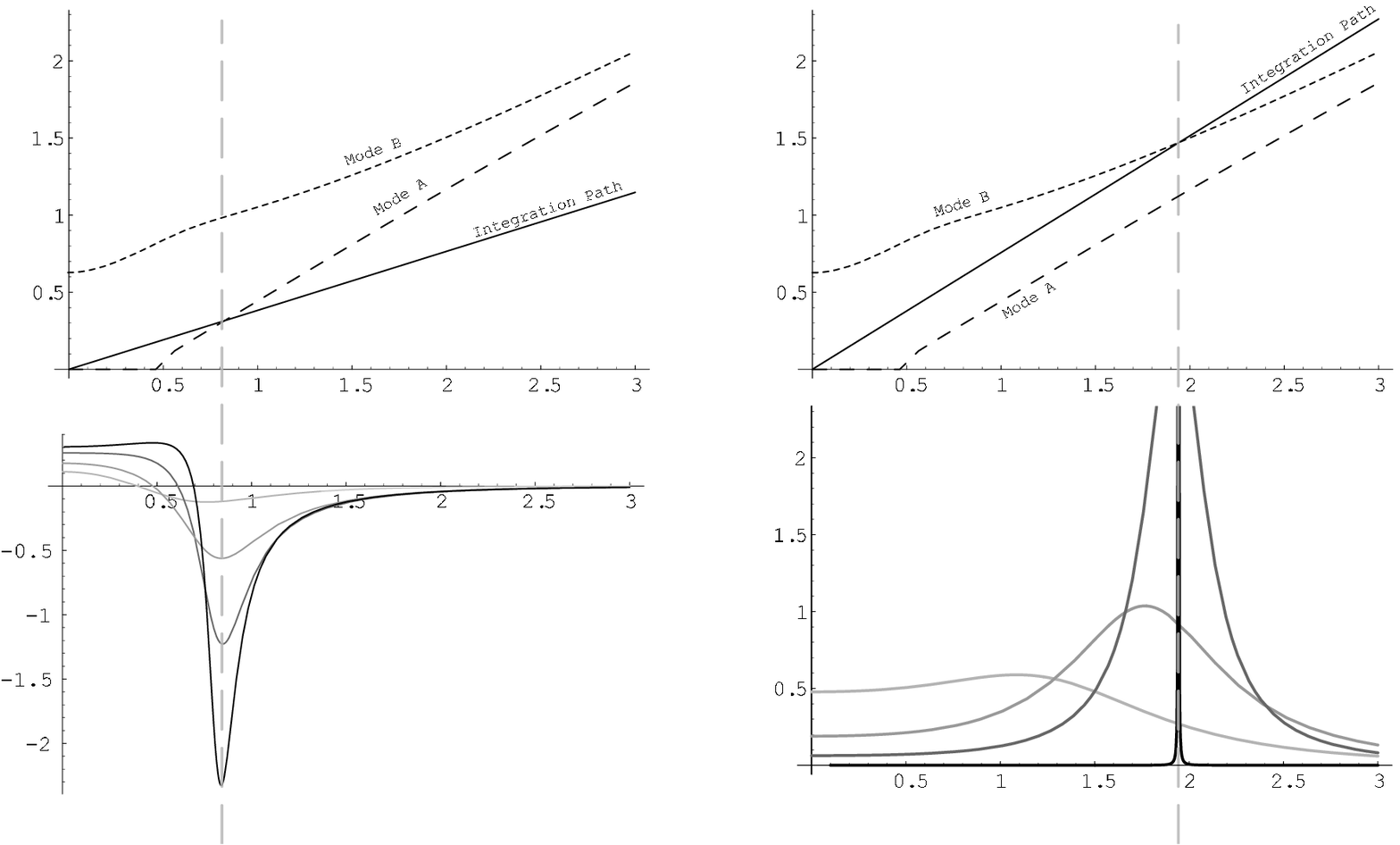}%

\setlength{\unitlength}{1cm}
\begin{picture}(15.5,0)
\put(0.0,2){\makebox(0,0){\begin{rotate}{90}%
\footnotesize $\sim -\;\frac{d^3 W}{dx \,d\theta_q\,dq^*}$
\end{rotate}}}
\put(8.3,2){\makebox(0,0){\begin{rotate}{90}%
\footnotesize $\sim -\;\frac{d^3 W}{dx \,d\theta_q\,dq^*}$
\end{rotate}}}
\put(7,3.7){\makebox(0,0){\footnotesize $q/m_D$}}
\put(7.1,5){\makebox(0,0){\footnotesize $q/m_D$}}
\put(15,0.8){\makebox(0,0){\footnotesize $q/m_D$}}
\put(15,5){\makebox(0,0){\footnotesize $q/m_D$}}
\put(0.5,9.2){\makebox(0,0){\footnotesize ${\omega\over m_D}$}}
\put(8.4,9.2){\makebox(0,0){\footnotesize ${\omega\over m_D}$}}
\put(4.3,10){\makebox(0,0){\footnotesize (a)}}
\put(12.1,10){\makebox(0,0){\footnotesize (b)}}
\end{picture}

\caption{In the top plots we show the integration 
paths corresponding to (a) $v=0.5$ and (b) $v=0.99$ as solid lines along  with 
the corresponding dispersion relations for the two infinite-$\xi$ unphysical 
modes (dashed and dotted lines labelled as ``Mode A'' and ``Mode B''). In the 
bottom plots we show the corresponding value 
of the integrand of Eq.~(\ref{elossReg2}) evaluated along the integration 
paths shown in the top plots. 
We also plot the integrand of Eq.~(\ref{elossReg2}) for $\xi=\{1000,100,10\}$
as decreasingly lighter gray lines in the bottom plot.
Note that in both (a) and (b) the integration paths in the top plots
should be understood to terminate at $q=q^*$.}
\label{fig:Intcomp1}
\end{figure}

As a further simplification we next consider the limit $\xi\rightarrow \infty$ 
for which the dispersion relations for the collective modes on the 
unphysical sheets are known for arbitrary angles $\theta_q$ of propagation 
\cite{Romatschke:2004jh}. This limit is convenient because it is possible to 
analytically determine the structure functions in this case, however, the 
general conclusions discussed below still apply when $\xi$ is finite. In the top 
plots of Fig.~\ref{fig:Intcomp1} we show the integration paths 
corresponding to (a) $v=0.5$ and (b) $v=0.99$ as solid lines along  with the 
corresponding dispersion relations for the two infinite-$\xi$ unphysical modes 
(dashed and dotted lines labelled as ``Mode A'' and ``Mode B''). In the bottom 
plots of Fig.~\ref{fig:Intcomp1} we show the corresponding value of the 
integrand of Eq.~(\ref{elossReg2}) evaluated along the integration paths shown 
in the top plots. Note that in both (a) and (b) the integration paths should be 
understood to terminate at $q=q^*$.

Let us discuss the case of $v=0.5$ first: as indicated by the vertical line in 
Fig.~\ref{fig:Intcomp1}(a), the extremal contribution to the energy loss occurs 
when the integration path comes closest to the mode A. However, it turns out the 
contribution from this mode is \emph{negative}, representing an inversion of the 
usual Landau-damping mechanism (we will discuss this issue later on). This 
situation is not limited to infinite anisotropies.  In the bottom plot of 
Fig.~\ref{fig:Intcomp1}(a) we show the integrand of Eq.~(\ref{elossReg2}) for 
$\xi=\{1000,100,10\}$ as decreasingly lighter gray lines.  As we can see the 
effect remains but becomes less as the anisotropy parameter is decreased.

In Fig.~\ref{fig:Intcomp1}(b) we show the same quantities but for 
$v=0.99$.\footnote{There is no ``regular'' contribution to the soft energy loss 
for $\xi\rightarrow \infty$,  since the structure functions have non-vanishing 
imaginary part only for $\hat\omega<\sin{\theta_q}$ 
\cite{Romatschke:2003yc,Romatschke:2004jh}, which together with $\hat\omega=v 
\cos{\theta_q}$ gives $v<\tan{\theta_q}$. However, since mode B moves onto the 
physical sheet for $\xi\rightarrow\infty$, the argument of the logarithm in 
Eq.~(\ref{Elosssoftfinal}), though real, changes sign there, picking up a phase 
which results in the $\delta$-function like peak in the lower plot of 
Fig.~\ref{fig:Intcomp1}(b) (black line).}  As indicated by the vertical 
line in Fig.~\ref{fig:Intcomp1}(b), the extremal contribution to the energy loss 
in this case occurs when the integration path comes closest to the mode B. 
However, in contrast, mode B contributions to the integrand are \emph{positive}. 
Again, this behavior persists at finite $\xi$ as shown by the corresponding 
plots of the integrand of Eq.~(\ref{elossReg2}) for $\xi=\{1000,100,10\}$ in 
Fig.~\ref{fig:Intcomp1}(b) (shown as increasingly lighter gray lines).  However,
the peak of the integrand for mode B contributions moves considerably more than
the mode A contributions as $\xi$ is decreased.

As shown by the light gray lines in the lower plots of Fig.~\ref{fig:Intcomp1}, 
the  unphysical A and B modes seem to have the same qualitative effect on the 
energy loss at finite $\xi$ as they do in the limit of $\xi\rightarrow \infty$. 
However, since the modes move farther away from the physical region as $\xi$ is 
decreased, it is clear that their influence becomes smaller, as already argued 
in Ref.~\cite{Romatschke:2004jh}. Nevertheless, it is interesting to note that 
even for small values of $\xi$, mode A is capable of driving the soft energy 
loss contribution negative for small velocities $v$. Indeed, it turns out that 
in the limit $v\rightarrow 0$, one finds that the soft energy loss is purely 
negative (independent of the angle), as long as $\xi>0$. 

This negative energy loss contribution is, however, \emph{not} due to the fact 
that a particle with sub-thermal velocities should receive energy from the 
medium, since our approximation was based on the assumption of an infinitely 
heavy quark (see the discussion about limitations in the next section). Rather, 
since mode A is the analytic continuation of the plasma instability present on 
the physical sheet, we interpret the gain of energy as being intimately related 
to the presence of such instabilities.  This is not a surprising finding given 
that if there are truly unstable modes in the system the energy for their growth 
must come from a mechanism which is also collective since it would be hard to imagine 
that a collisional mechanism could ever be fast enough.  

The mode A contributions to the soft energy ``loss'' represent such a collective 
mechanism for the efficient transfer of energy from hard to soft scales.  As we 
will show below the energy gain is focused in such a way that particles with 
small velocities are accelerated out-of-plane thereby reducing the anisotropy. 
In closing we note that for angles different than $\theta_v=0$, the connection 
between the  modes on the neighboring unphysical sheets and the integrand of the 
energy loss remains but it is more difficult to disentangle since there is 
also an averaging over the azimuthal angle as well.

\section{Heavy Quark Energy Loss}
\label{QCDEloss}

The collisional energy loss of a heavy quark in an anisotropic quark-gluon
plasma to leading order in the coupling is given by adding the contributions
Eq.~(\ref{Elosssoftfinal}) and Eq.~(\ref{myelosshard3})
\beq
\left(\frac{dW}{dx}\right)=
\left(\frac{dW}{dx}\right)_{\rm soft}+%
\left(\frac{dW}{dx}\right)_{\rm hard}.
\label{Elosstotal}
\eeq
\checked{mp}
In all results presented below we will assume $N_c=3$ and $N_f=2$.

\subsection{Isotropic results}

\begin{figure}
\hfill
\begin{minipage}[t]{.45\linewidth}
\includegraphics[width=0.9\linewidth]{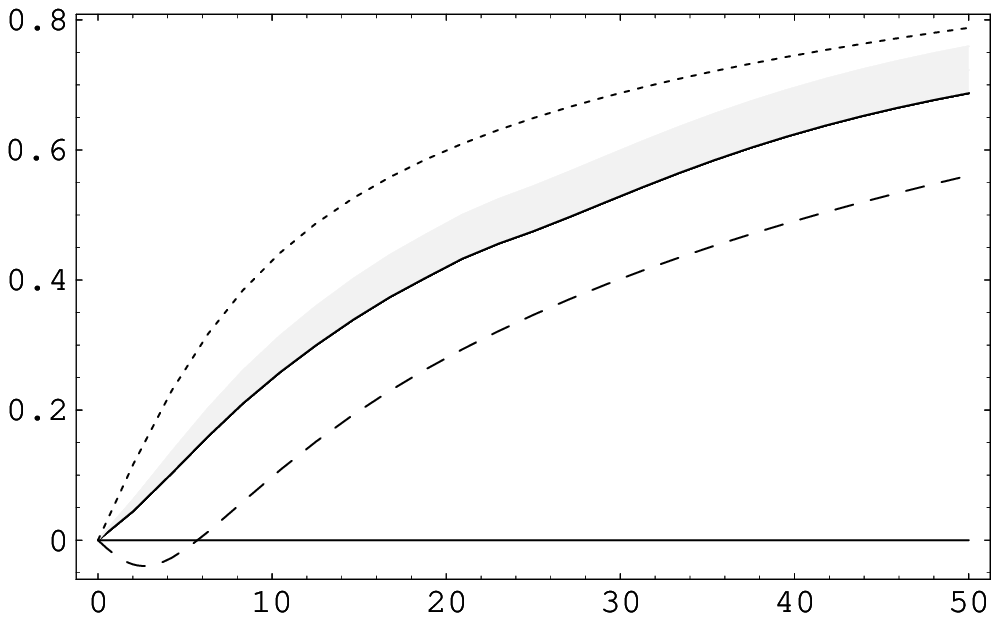}
\setlength{\unitlength}{1cm}
\begin{picture}(9,0)
\put(-0.1,1){\makebox(0,0){\begin{rotate}{90}%
\footnotesize
$-\left(\frac{dW}{dx}\right)_{\rm iso}%
\;[{\rm GeV/fm}]$ 
\end{rotate}}}
\put(4,0.25){\makebox(0,0){\footnotesize $p\;[{\rm GeV}]$}}
\put(4.1,4.9){\makebox(0,0){\rm (a)}}
\end{picture}
\end{minipage} \hfill
\hfill
\begin{minipage}[t]{.45\linewidth}
\includegraphics[width=0.9\linewidth]{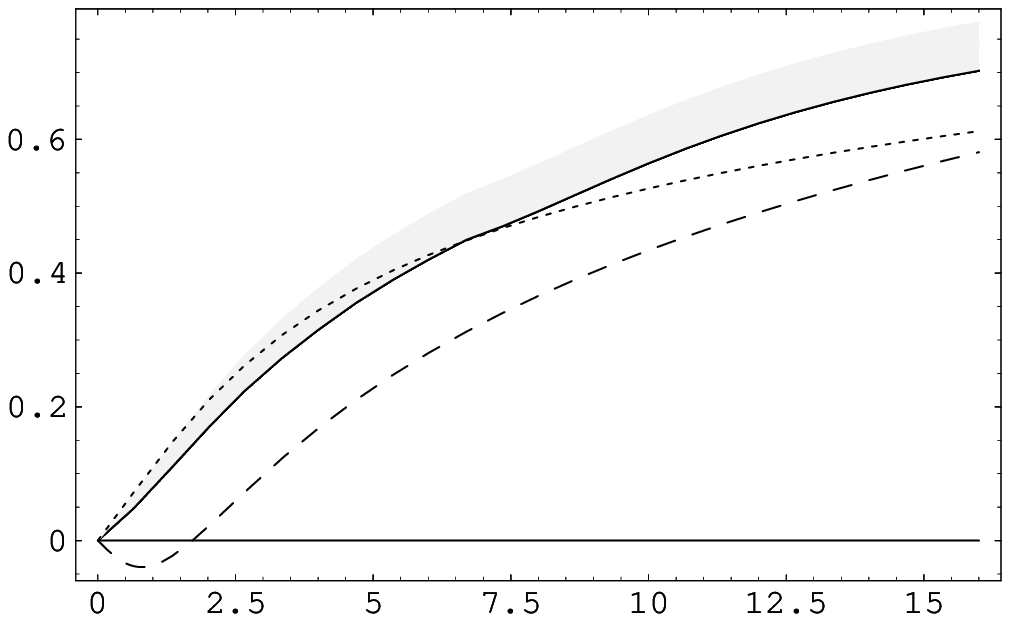}
\setlength{\unitlength}{1cm}
\begin{picture}(9,0)
\put(-0.1,1){\makebox(0,0){\begin{rotate}{90}%
\footnotesize $-\left(\frac{dW}{dx}\right)_{\rm iso}%
\;[{\rm GeV/fm}]$ 
\end{rotate}}}
\put(4,0.25){\makebox(0,0){\footnotesize $p\;[{\rm GeV}]$}}
\put(4.1,4.9){\makebox(0,0){\rm (b)}}
\end{picture}
\vspace{-1cm}
\end{minipage} \hfill
\caption{Isotropic
energy loss of a bottom (a) and charm quark (b) as a function
of momentum $p$ for $\alpha_s=0.3$. Shown are the respective
results from Bjorken \cite{Bjorken:1982tu} 
(dotted line) and Braaten and Thoma \cite{Braaten:1991we} (dashed line)
as well as the evaluation
of Eq.~(\ref{Elosstotal}) (full line) with a variation of $q^*$ (gray band).}
\label{fig:QCDELq}
\end{figure}

In order to compare with existing calculations we first consider the isotropic 
limit ($\xi=0$).  In this case the collisional energy loss is a function of the 
strong coupling $\alpha_s$, the particle velocity $v$, the temperature $T$, and 
the momentum separation scale $q^*$. The $q^*$ dependence of the result is found 
to become weak for small values of the coupling limit $\alpha_s$ (similar to 
what has been found in QED \cite{Romatschke:2003vc}), corresponding to the 
original result by Braaten and Thoma \cite{Braaten:1991we}; for larger values of 
the coupling one fixes $q^*=q^{\rm pms}$ using the principle of minimum 
sensitivity
\beq
\frac{d }{d q^*} \left. \left(\frac{dW}{dx}\right) %
\right|_{q^*=q^*_{\rm pms}}=0 \, .
\label{PMScond}
\eeq 
\checked{mp}
Note that in this case $-\left. \left(dW/dx\right) %
\right|_{q^*=q^*_{\rm pms}}$ always serves as a lower bound on the result for 
the energy loss. To get an estimate of how strongly the result depends on this 
special value of $q^{*}$, one can e.g. vary $q^*_{\rm pms}$ by a certain factor 
$c_{q^*}$ and evaluate the energy loss at $q^*_{\rm pms} c_{q^*}$ and $q^*_{\rm 
pms}/c_{q^*}$. This factor $c_{q^*}$ is in principle arbitrary; here we fix it 
to be $c_{q^*}=2$ to be consistent with what has been done in QED 
\cite{Romatschke:2003vc}.

In Fig.~\ref{fig:QCDELq} various results for the heavy quark energy loss in 
an isotropic quark-gluon plasma are compared: shown are the results from Bjorken 
\cite{Bjorken:1982tu}, Braaten and Thoma \cite{Braaten:1991we}, as well as 
Eq.~(\ref{Elosstotal}) with $\xi=0$ at $q^*=q^*_{\rm pms}$ together with its 
variation using $c_{q^*}=2$. Both plots assume $\alpha_s=0.3$. 
Fig.~\ref{fig:QCDELq}(a) shows the energy loss of a bottom quark with mass 
$M_{Q} = 5$~GeV while in Fig.~\ref{fig:QCDELq}(b) the energy loss of a charm 
quark with mass $M_{Q} = 1.5$~GeV is plotted, both as a function of their 
momenta
\beq
p=\frac{v M_Q}{\sqrt{1-v^2}} \; .
\label{pvrel}
\eeq
\checked{mp}

In closing we would like to emphasize that our result is not only compatible 
with previous results on isotropic collisional energy loss but, in fact, 
represents the most complete calculation of this quantity to date.  In contrast 
to the result of Bjorken \cite{Bjorken:1982tu} it gives the correct leading-order
result in the weak-coupling limit as does the result of Braaten and Thoma 
\cite{Braaten:1991we}; however, as noted before, at realistic values of the 
coupling and small momentum the Braaten and Thoma result breaks down completely. 
In the isotropic case our result is positive definite as it should be for an 
infinitely massive quark. In addition the use of a principle-of-minimal 
sensitivity to fix the scale gives the added benefit of being able to place 
theoretical error bars on the results obtained.

\subsection{Limitations}

Since Eq.~(\ref{Elosstotal}) has been derived assuming an infinitely heavy quark 
we have to determine what will happen if we instead have only a heavy quark like 
a charm or bottom quark.  The chief consequence is that for sub-thermal 
velocities defined by $v\,\raisebox{-1mm}{$\stackrel{<}{\sim}$}\,\sqrt{T/M_Q}$ 
the approximation breaks down because quarks with those velocities will gain 
energy from collisions with other particles in the heat bath instead of losing
energy. Below this threshold the energy loss would then turn into energy gain. 
A semi-quantitative estimate of the velocity where the isotropic energy loss 
becomes negative has been performed in Ref.~\cite{Braaten:1991we} by repeating 
the above calculation in the limit $v\rightarrow 0$ and for weak couplings, 
finding $v\,\raisebox{- 1mm}{$\stackrel{<}{\sim}$}\,\sqrt{3 T/M_Q}$.  Using 
$T=250$~MeV and the masses above, this velocity corresponds to $p=2.1$~GeV and 
$p=1.5$~GeV for bottom and charm, respectively. Similarly, 
Eq.~(\ref{Elosstotal}) also breaks down for ultrarelativistic energies $E\gg 
M_Q^2/T$, with the cross-over energy having been determined to be $E_{\rm 
cross}\simeq 1.8\,M_Q^2/T$ (corresponding to $v>0.995$ for both charm and bottom 
quarks). 

In contrast the Braaten and Thoma results shown in Fig.~\ref{fig:QCDELq} turn 
negative at a momentum of $p\simeq 5.7$~GeV for the bottom and $p\simeq 
1.71$~GeV for the charm quark.  This behavior is not due to the physical reason 
discussed above but rather due to a failure of the extrapolation from the weak-coupling 
limit to realistic couplings \cite{Braaten:1991we}. For the unexpanded 
isotropic result Eq.~(\ref{Elosstotal}), this unphysical behavior does not occur 
(it is always positive) and one can therefore expect it to be valid for 
velocities down to the original estimate $v\sim \sqrt{3 T/M_Q}$.  In anisotropic 
systems, however, this rule of thumb no longer applies and negative energy loss 
does not require the heavy quark to have sub-thermal velocity.  Instead, as 
already noted by Lifshitz and Pitaevskii \cite{Pitae:1981}, this type of energy 
gain is connected to the instabilities of the system, as we have argued in the 
previous section.

\begin{figure}[t]
\begin{center}
\includegraphics[width=3.5cm]{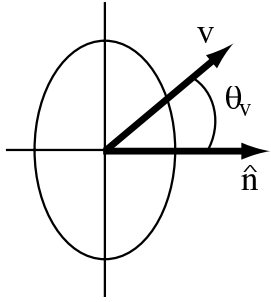}
\caption{Sketch of the vectors relevant to the directional dependence of the energy loss: 
$\bf \hat{n}$ is the direction of the anisotropy, ${\bf v}$ the velocity 
of the quark, and $\theta_v$ the angle between these.  We have also plotted
an isocontour of a $\xi>0$ distribution function for comparison.}
\label{fig:ELsquas}
\end{center}
\end{figure}

Before moving on to the anisotropic results let us briefly comment on the 
variation estimate based on setting $c_{q^*}=2$: while for QED this represents a 
very conservative choice on the variations since $m_D\ll q^*_{pms}\ll 2 \pi T$, 
in QCD one encounters situations with $q^*_{pms}>2\pi T$ or $q^*_{pms}<m_D$ for 
realistic couplings like $\alpha_s=0.3$ since $m_D$ and $2\pi T$ are no longer 
well separated. Strictly speaking, the method we are using is breaking down at 
realistic couplings; nevertheless, by using variations with $c_{q^*}=2$ 
we can hope to obtain reasonable estimates of  the leading-order 
results when the difference between these variations is not too large. Finally, 
it should be noted that no estimate of the next-to-leading order (NLO) radiative 
corrections to the energy loss has been made here. Therefore, it should be kept 
in mind that for QCD with large realistic coupling the inclusion of these NLO 
corrections might give energy loss results that are not covered by the 
variations of $q^*$ in the leading-order result, so these variations should be 
interpreted with care.

\subsection{Anisotropic results}

\begin{figure}
\hfill
\begin{minipage}[t]{.45\linewidth}
\includegraphics[width=0.9\linewidth]{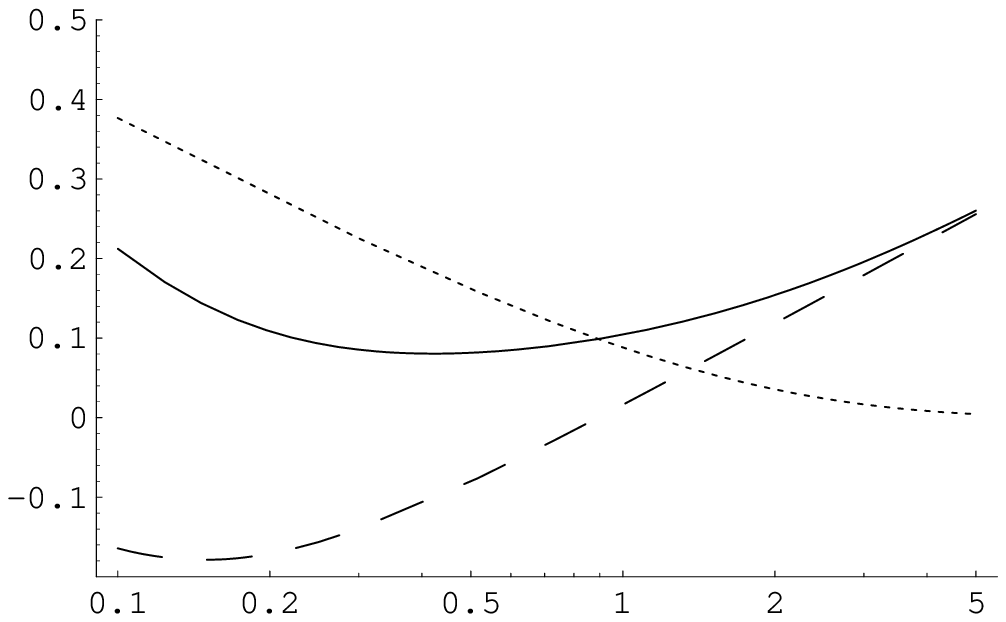}
\setlength{\unitlength}{1cm}
\begin{picture}(9,0)
\put(-0.1,1.2){\makebox(0,0){\begin{rotate}{90}%
\footnotesize $-\left(\frac{dW}{dx}\right)/\left(\frac{8\pi^2\alpha_s^2T^2}{3}\right)$ 
\end{rotate}}}
\put(4,0.25){\makebox(0,0){\footnotesize $q^*/T$}}
\end{picture}
\vspace{-1cm}
\caption{Anisotropic energy loss for $\xi=10.0$, 
$\alpha_s=1\times10^{-4}$, $v=0.1$, and $\theta_v=0$ as a function of $q^*$.  Dotted, dashed,
and solid lines are the hard contribution, the soft contribution, and
their sum, respectively.}
\label{fig:QCDas0001}
\end{minipage} \hfill
\hfill
\begin{minipage}[t]{.45\linewidth}
\includegraphics[width=0.9\linewidth]{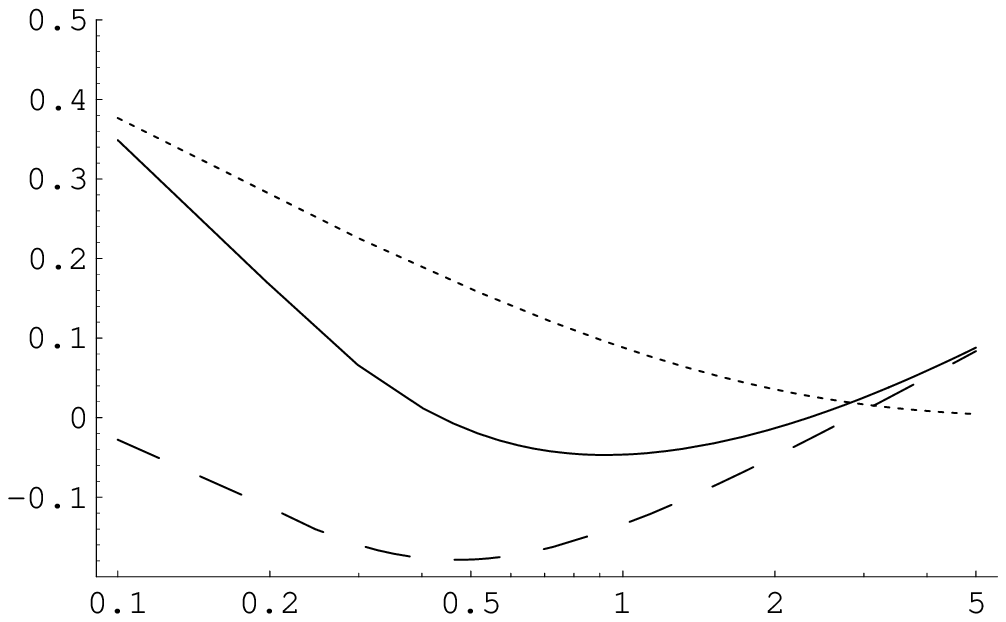}
\setlength{\unitlength}{1cm}
\begin{picture}(9,0)
\put(-0.1,1.2){\makebox(0,0){\begin{rotate}{90}%
\footnotesize $-\left(\frac{dW}{dx}\right)/\left(\frac{8\pi^2\alpha_s^2T^2}{3}\right)$ 
\end{rotate}}}
\put(4,0.25){\makebox(0,0){\footnotesize $q^*/T$}}
\end{picture}
\vspace{-1cm}
\caption{Anisotropic energy loss for $\xi=10.0$, 
$\alpha_s=1\times10^{-3}$, $v=0.1$, and $\theta_v=0$ as a function of $q^*$.  Dotted, dashed,
and solid lines are the hard contribution, the soft contribution, and
their sum, respectively.}
\label{fig:QCDas001}
\end{minipage} \hfill
\end{figure}

\begin{figure}
\hfill
\begin{minipage}[t]{.45\linewidth}
\includegraphics[width=0.9\linewidth]{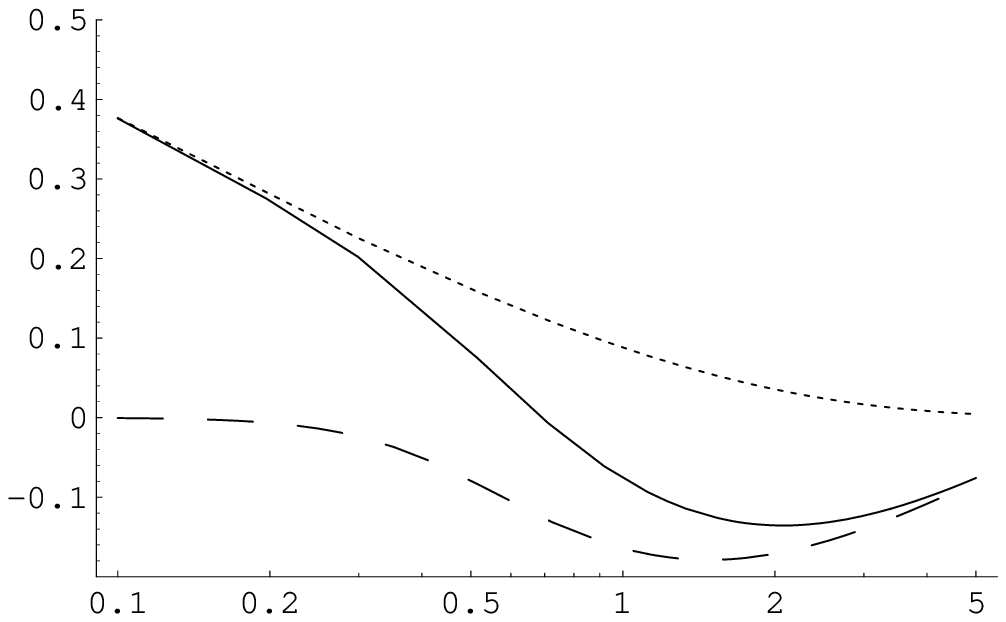}
\setlength{\unitlength}{1cm}
\begin{picture}(9,0)
\put(-0.1,1.2){\makebox(0,0){\begin{rotate}{90}%
\footnotesize $-\left(\frac{dW}{dx}\right)/\left(\frac{8\pi^2\alpha_s^2T^2}{3}\right)$ 
\end{rotate}}}
\put(4,0.25){\makebox(0,0){\footnotesize $q^*/T$}}
\end{picture}
\vspace{-1cm}
\caption{Anisotropic energy loss for $\xi=10.0$, 
$\alpha_s=0.01$, $v=0.1$, and $\theta_v=0$ as a function of $q^*$.  Dotted, dashed,
and solid lines are the hard contribution, the soft contribution, and
their sum, respectively.}
\label{fig:QCDas01}
\end{minipage} \hfill
\hfill
\begin{minipage}[t]{.45\linewidth}
\includegraphics[width=0.9\linewidth]{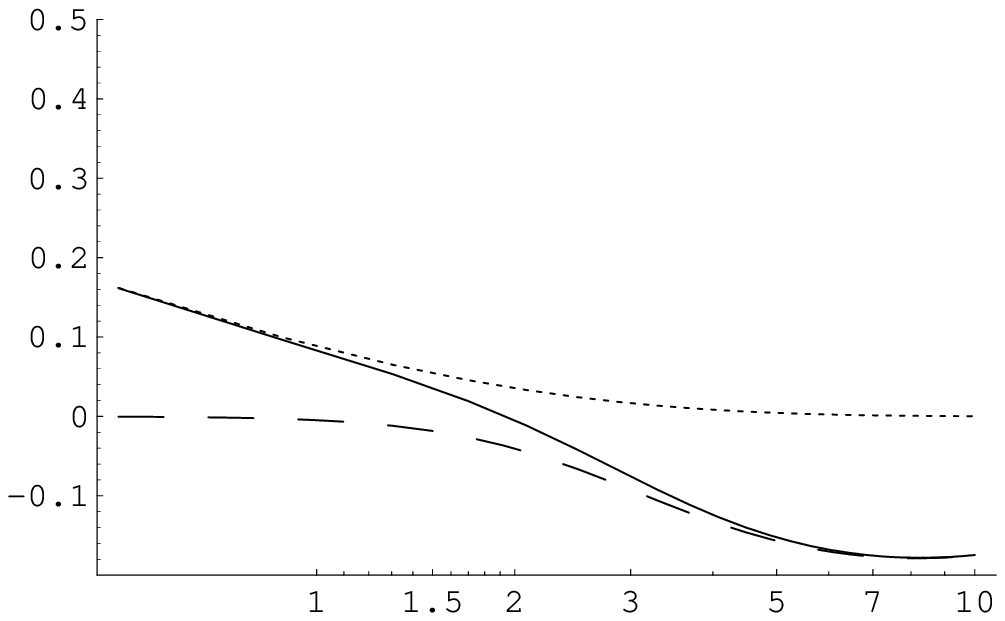}
\setlength{\unitlength}{1cm}
\begin{picture}(9,0)
\put(-0.1,1.2){\makebox(0,0){\begin{rotate}{90}%
\footnotesize $-\left(\frac{dW}{dx}\right)/\left(\frac{8\pi^2\alpha_s^2T^2}{3}\right)$ 
\end{rotate}}}
\put(4,0.25){\makebox(0,0){\footnotesize $q^*/T$}}
\end{picture}
\vspace{-1cm}
\caption{Anisotropic energy loss for $\xi=10.0$, 
$\alpha_s=0.3$, $v=0.1$, and $\theta_v=0$ as a function of $q^*$.  Dotted, dashed,
and solid lines are the hard contribution, the soft contribution, and
their sum, respectively.}
\label{fig:QCDas3}
\end{minipage} \hfill
\end{figure}

For an anisotropic system with (gluon and quark) anisotropy strength $\xi$, the 
collisional energy loss result Eq.~(\ref{Elosstotal}) depends on $q^*$, $T$, and 
$v$ as was the case for the isotropic energy loss, but in general also on the 
angle of the particle propagation with respect to the anisotropy vector 
$\cos\theta_v={\bf \hat{v}\cdot \hat{n}}$ (sketched in Fig.~\ref{fig:ELsquas}). 
As in the isotropic case the total result is obtained by combining the soft and 
hard contributions to the energy loss.  In Figs.~\ref{fig:QCDas0001}-\ref{fig:QCDas3} 
we plot the hard and soft contributions to the energy loss 
together with their sum as a function of $q^*/T$ for $\alpha_s=\{1\times10^{-4}, 
1\times10^{-3},0.01,0.3\}$ corresponding to $m_D/T=\{0.041,0.13,0.41,2.2\}$.  
As can be seen from these Figures the soft contribution 
(dashed line) is always negative over some range of $q^*/T$.   For small 
couplings (see Fig.~\ref{fig:QCDas0001}) this negative contribution is in a 
region where there is always a larger positive hard contribution.  However, as 
the coupling constant is increased we see that the range of $q^*/T$ over which 
the soft contribution is negative moves to higher scales and eventually 
dominates the sum of the soft and hard contributions over a range of $q^*$ (see 
Fig.~\ref{fig:QCDas3}). 

For all $\xi$ we find that the soft energy loss has a minimum at some value of 
$q^*$.  For small $v$ this minimum can even be negative as explicitly 
demonstrated in Figs.~\ref{fig:QCDas01} and \ref{fig:QCDas3}.   Considering 
again $\theta_v=0$ for guidance we notice that the limits on the integration in 
Eq.~(\ref{elossReg2}) correspond to a triangular region defined by $0 \leq 
\omega \leq qv$ and $q \leq q^*$. From this we observe that when $q^*$ is small 
we only receive contributions from mode A which are predominantly negative.  As 
$q^*$ is increased, however, we start to receive large positive contributions 
from mode B. Defining $q^*_c$ as the point where the upper boundary defined by 
$\omega = q v$ hits mode B we obtain $q^*_c = \omega_B/v$.  In order to estimate 
the value of $q_c^*$ we again consider the limit $\xi\rightarrow \infty$.  The 
dispersion relation for mode B at $\theta_q=0$ and $\xi=\infty$ is known and is 
$q$-independent \cite{Romatschke:2004jh}
\beq
\omega_B^2=\frac{\pi}{4} m_D^2 \, .
\eeq
\checked{mp}
Using this we obtain
\beq
q_c^*=\sqrt{\frac{\pi}{4}} \frac{m_D}{v} \, .
\eeq
\checked{mp}
For larger values of $q^*$, the negative contributions also
become larger, but there is also now a strongly positive contribution from mode B
(even for nonzero $\theta_q$), 
so that the overall result turns out to
be more positive than at $q^*_c$. On the other hand, for $q^*<q^*_c$,
the main negative contributions become smaller since the ``region''
they are integrated over has shrunk. Therefore, the minimum of the
soft contribution to the energy loss for $\xi\rightarrow \infty$
at $\theta_v=0$ is obtained around $q^*_c$.  This can be used as a rough
guideline for the location of this minimum also for large but 
finite $\xi$, however, as $\xi$ is decreased $q_c^*$ moves to lower
values of $q^*$ and the depth of the minimum decreases.

Returning to the results for the total anisotropic energy loss (hard plus soft) 
we again fix the scale $q^*$ by the application of the principle of minimal 
sensitivity (\ref{PMScond}) as in the isotropic case; however, in order to keep 
the figures presented as understandable as possible we do not show the 
corresponding variation of $q^*$ by $c_{q^*}=2$. Instead we note that the 
sensitivity of the anisotropic results to this scale is similar to the isotropic 
variation indicated by the gray band in Fig.~\ref{fig:QCDELq}, however, there is 
an increase in the magnitude of the variation as $\xi$ is increased. For 
example, at $\xi=\infty$ the corresponding variation is approximately three 
times the corresponding isotropic variation.

\begin{figure}
\hfill
\begin{minipage}[t]{.45\linewidth}
\includegraphics[width=0.9\linewidth]{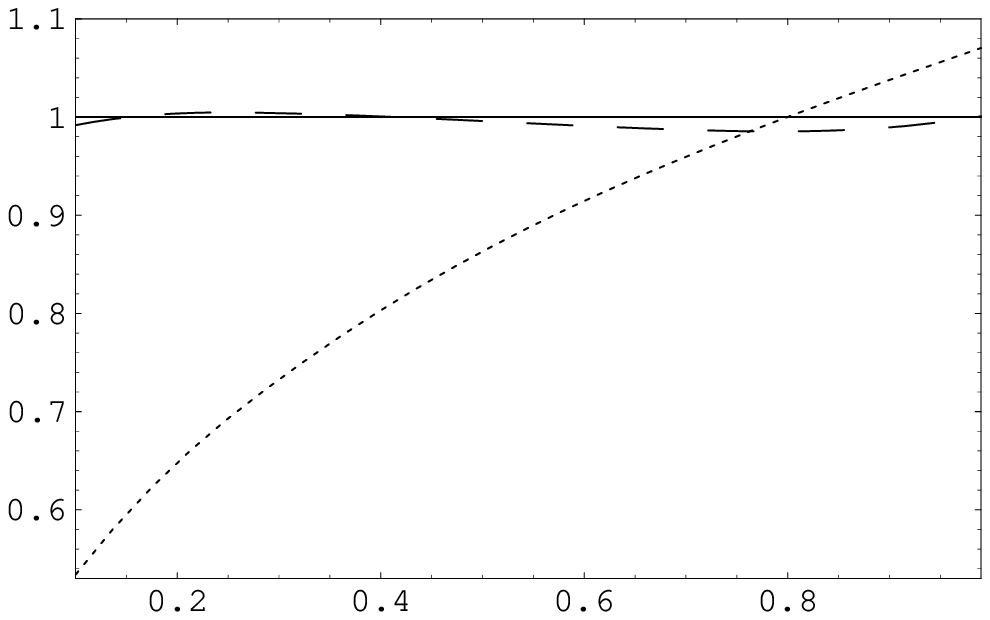}
\setlength{\unitlength}{1cm}
\begin{picture}(9,0)
\put(-0.1,1.5){\makebox(0,0){\begin{rotate}{90}%
\footnotesize
$\left(\frac{dW}{dx}\right)/\left(\frac{dW}{dx}\right)_{\rm iso}$
\end{rotate}}}
\put(4,0.25){\makebox(0,0){\footnotesize $v$}}
\end{picture}
\vspace{-1cm}
\caption{Anisotropic energy loss for $\xi=1.0$, 
$\alpha_s=0.3$ and $\theta_v=0$ (dotted line)
compared to $\theta=\pi/2$ (dashed line) as a function of velocity 
(normalized to the isotropic result, full line).}
\label{fig:QCDnxi1}
\end{minipage} \hfill
\hfill
\begin{minipage}[t]{.45\linewidth}
\includegraphics[width=0.9\linewidth]{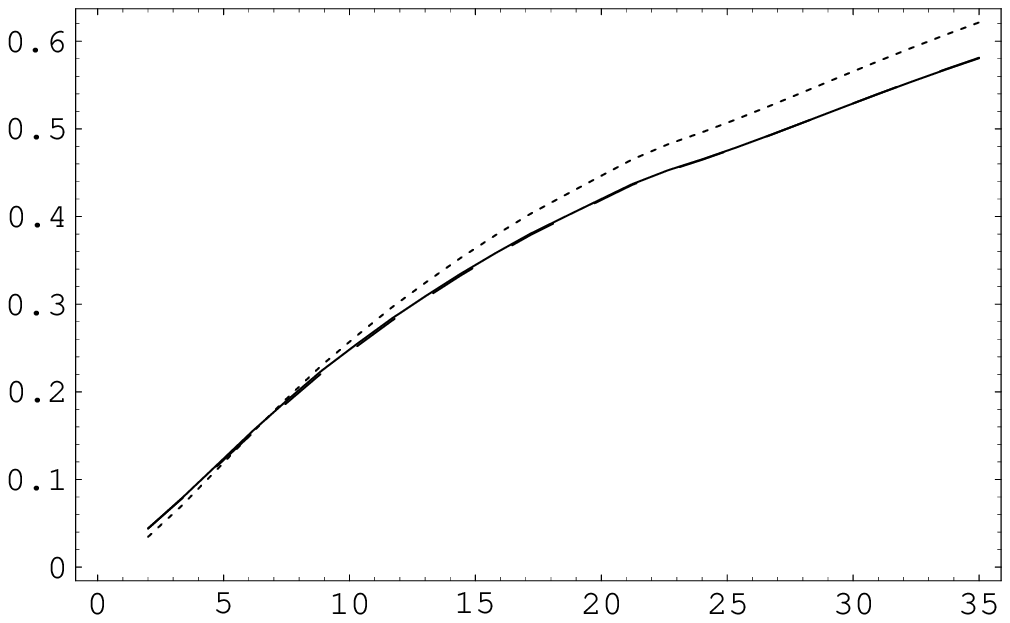}
\setlength{\unitlength}{1cm}
\begin{picture}(9,0)
\put(-0.1,1.2){\makebox(0,0){\begin{rotate}{90}%
\footnotesize $-\left(\frac{dW}{dx}\right)%
\;[{\rm GeV/fm}]$ 
\end{rotate}}}
\put(4,0.25){\makebox(0,0){\footnotesize $p\;[{\rm GeV}]$}}
\end{picture}
\vspace{-1cm}
\caption{Anisotropic energy loss for a bottom quark for $\xi=1.0$, 
$\alpha_s=0.3$ and $\theta_v=0$ (dotted line), 
$\theta_v=\pi/2$ (dashed line) compared to the isotropic result (full line).}
\label{fig:QCDxi1}
\end{minipage} \hfill
\end{figure}

In Fig.~\ref{fig:QCDnxi1} we show the energy loss Eq.~(\ref{Elosstotal}) for a 
heavy quark propagating parallel to the anisotropy direction ($\theta_v=0$) and 
perpendicular to the anisotropy direction ($\theta_v=\pi/2$) as a function of 
the heavy quark velocity $v$ with $\alpha_s=0.3$ and $\xi=1$.  The 
results shown are normalized to the isotropic energy loss. As can be seen from this 
Figure for small velocities the energy loss is larger at $\theta_v=\pi/2$
while for large velocities it is larger at $\theta_v=0$.  This is similar to
the behavior found for a QED plasma \cite{Romatschke:2003vc}.  

In Fig.~\ref{fig:QCDxi1} we show the corresponding energy loss for a 
bottom quark with $M_Q = 5$~GeV using the relation between velocity and momentum 
given in Eq.~(\ref{pvrel}).  We have not normalized this result as before but 
instead include the corresponding isotropic result as a solid line for 
comparison.  As can be seen from this Figure for $\xi=1$ and $\alpha_s=0.3$ the 
energy loss of a bottom quark with $\theta_v=\pi/2$ is almost indistinguishable 
from the isotropic result while the energy loss with $\theta_v=0$ is larger than 
the isotropic result for $p \,\raisebox{-1mm}{$\stackrel{>}{\sim}$} \,6$~GeV. 
For example, for a 20~GeV bottom quark we find that the energy loss at 
$\theta_v=0$ is approximately 10\% higher than the isotropic result.

It should be noted that for very small velocities, the energy loss for the angle 
$\theta_v=0$ becomes negative because of the presence of modes on the 
unphysical sheet as discussed in Sec.~\ref{sec:umodes}.  However, for $\xi=1.0$ 
this occurs at very small velocities where the heavy quark approximation used is 
invalid in any case so that within the valid range of velocities ($v\geq0.39$ 
and $v\geq0.71$ for bottom and charm quarks, respectively) the energy loss 
is always positive.

\begin{figure}
\hfill
\begin{minipage}[t]{.45\linewidth}
\includegraphics[width=0.9\linewidth]{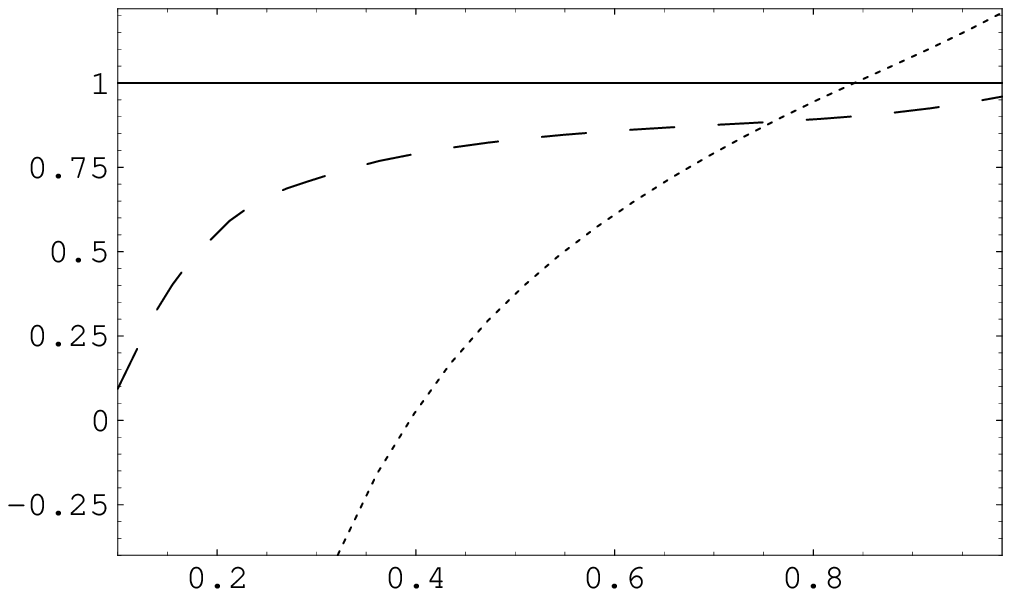}
\setlength{\unitlength}{1cm}
\begin{picture}(9,0)
\put(-0.1,1.5){\makebox(0,0){\begin{rotate}{90}%
\footnotesize
$\left(\frac{dW}{dx}\right)/\left(\frac{dW}{dx}\right)_{\rm iso}$
\end{rotate}}}
\put(4,0.25){\makebox(0,0){\footnotesize $v$}}
\end{picture}
\vspace{-1cm}
\caption{Anisotropic energy loss for $\xi=10.0$, 
$\alpha_s=0.3$ and $\theta_v=0$ (dotted line)
compared to $\theta=\pi/2$ (dashed line) as a function of velocity 
(normalized to the isotropic result, full line).}
\label{fig:QCDnxi10}
\end{minipage} \hfill
\hfill
\begin{minipage}[t]{.45\linewidth}
\includegraphics[width=0.9\linewidth]{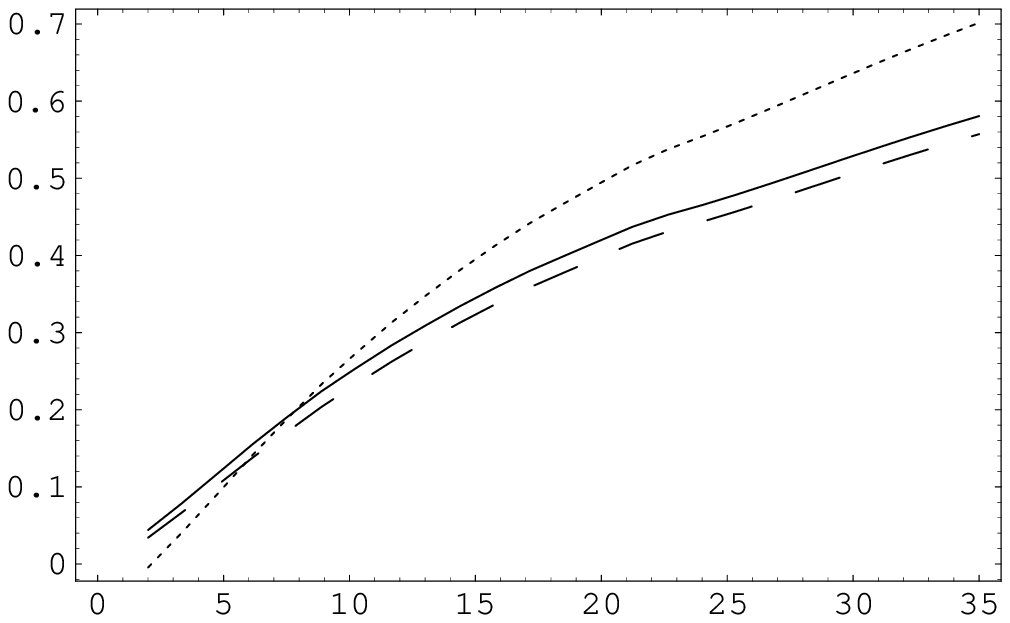}
\setlength{\unitlength}{1cm}
\begin{picture}(9,0)
\put(-0.1,1.2){\makebox(0,0){\begin{rotate}{90}%
\footnotesize $-\left(\frac{dW}{dx}\right)%
\;[{\rm GeV/fm}]$ 
\end{rotate}}}
\put(4,0.25){\makebox(0,0){\footnotesize $p\;[{\rm GeV}]$}}
\end{picture}
\vspace{-1cm}
\caption{Anisotropic energy loss for a bottom quark for $\xi=10.0$, 
$\alpha_s=0.3$ and $\theta_v=0$ (dotted line), 
$\theta_v=\pi/2$ (dashed line) compared to the isotropic result (full line).}
\label{fig:QCDxi10}
\end{minipage} \hfill
\end{figure}

In Figs.~\ref{fig:QCDnxi10} and \ref{fig:QCDxi10} we present the same plots but for 
$\xi=10$. As can be seen from Fig.~\ref{fig:QCDnxi10} for small velocities the 
energy loss is larger at $\theta_v=\pi/2$ while for large velocities it is 
larger at $\theta_v=0$ similar to the situation at $\xi=1$. From 
Fig.~\ref{fig:QCDxi10} we see that the bottom quark energy loss at 
$\theta_v=\pi/2$ is less than the isotropic result while the energy loss at 
$\theta_v=0$ is larger than the isotropic result for $p\,\raisebox{-
1mm}{$\stackrel{>}{\sim}$}\,8$~GeV. For example, for a 20~GeV bottom quark we find 
that the energy loss for $\theta_v=\pi/2$ is approximately 10\% lower than the 
isotropic result and for $\theta_v=0$ the energy loss is approximately 20\% 
higher than the isotropic result.  Again for velocities less than $v\sim0.4$ the 
energy loss at $\theta_v=0$ becomes negative; however, now the velocity where it 
becomes negative is on the border of the applicability of the heavy quark 
approximation for a bottom quark given in the previous paragraph. One is 
therefore lead to wonder what happens if we continue to increase the anisotropy.

\begin{figure}
\hfill
\begin{minipage}[t]{.45\linewidth}
\includegraphics[width=0.9\linewidth]{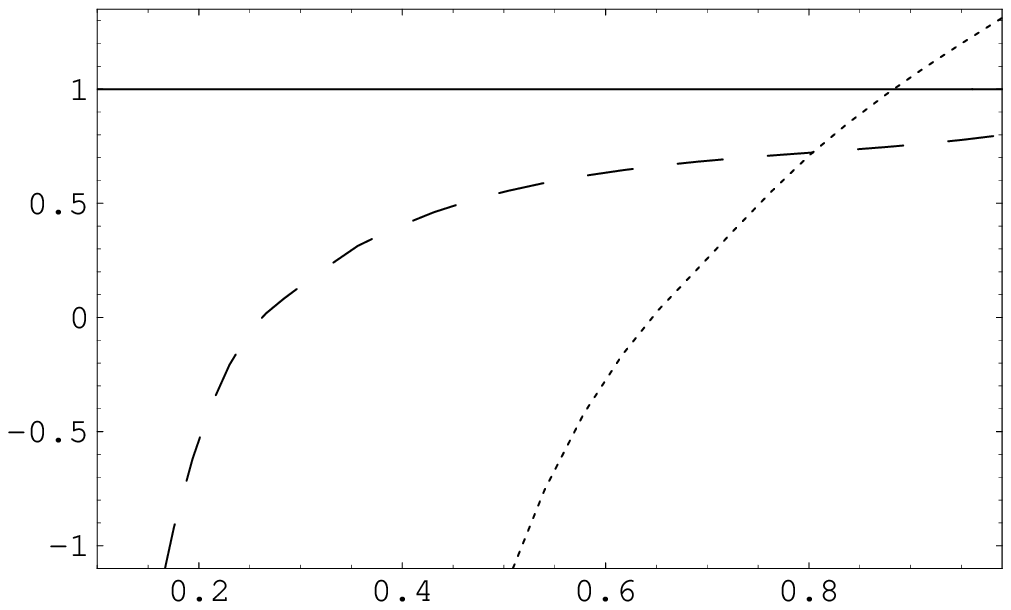}
\setlength{\unitlength}{1cm}
\begin{picture}(9,0)
\put(-0.1,1.5){\makebox(0,0){\begin{rotate}{90}%
\footnotesize 
$\left(\frac{dW}{dx}\right)/\left(\frac{dW}{dx}\right)_{\rm iso}$
\end{rotate}}}
\put(4,0.25){\makebox(0,0){\footnotesize $v$}}
\end{picture}
\vspace{-1cm}
\caption{Anisotropic energy loss for $\xi\rightarrow \infty$, 
$\alpha_s=0.3$ and $\theta_v=0$ (dotted line)
compared to $\theta_v=\pi/2$ (dashed line) as a function of velocity 
(normalized to the isotropic result, full line).}
\label{fig:QCDnLX}
\end{minipage} \hfill
\hfill
\begin{minipage}[t]{.45\linewidth}
\includegraphics[width=0.9\linewidth]{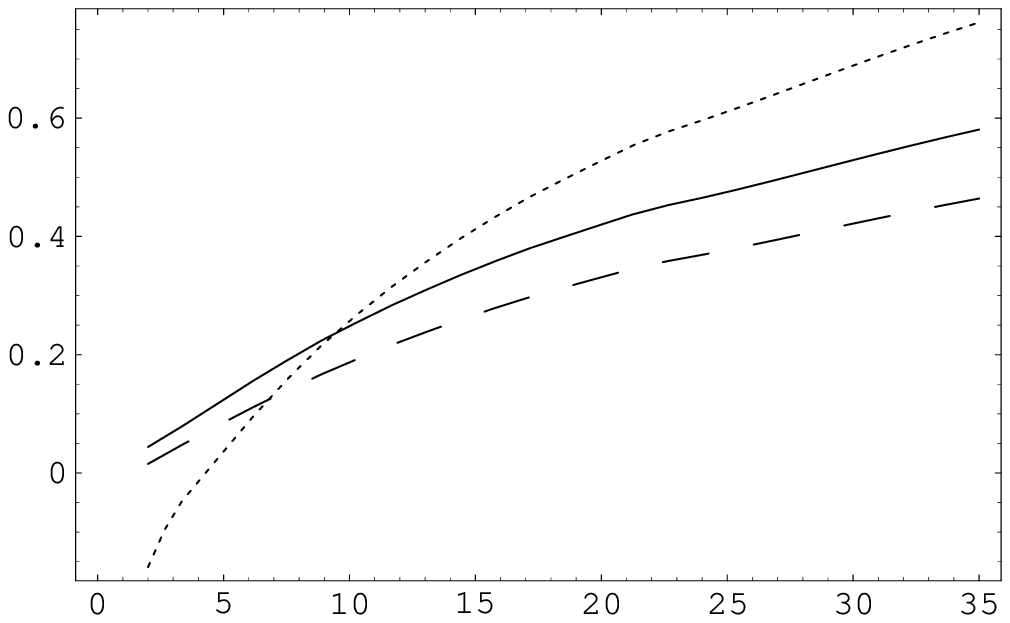}
\setlength{\unitlength}{1cm}
\begin{picture}(9,0)
\put(-0.1,1.2){\makebox(0,0){\begin{rotate}{90}%
\footnotesize $-\left(\frac{dW}{dx}\right)%
\;[{\rm GeV/fm}]$ 
\end{rotate}}}
\put(4,0.25){\makebox(0,0){\footnotesize $p\;[{\rm GeV}]$}}
\end{picture}
\vspace{-1cm}
\caption{Anisotropic energy loss for a bottom quark for 
$\xi\rightarrow \infty$, 
$\alpha_s=0.3$ and $\theta_v=0$ (dotted line), 
$\theta_v=\pi/2$ (dashed line) compared to the isotropic result (full line).
}
\label{fig:QCDLX}
\end{minipage} \hfill
\end{figure}

\subsection{Infinite anisotropy limit}

As Figs.~\ref{fig:QCDxi1} and \ref{fig:QCDxi10} demonstrate, as the anisotropy strength
$\xi$ is increased the directional dependence of the heavy quark energy loss 
increases.  Additionally we found that the velocity at which the energy loss
becomes negative increases as $\xi$ is increased.  It would be very instructive
to then consider the limit of $\xi\rightarrow\infty$ to see the limiting behavior
of the energy loss.

In Figs.~\ref{fig:QCDnLX} and \ref{fig:QCDLX} we present the same plots but for 
$\xi=\infty$. As can be seen from Fig.~\ref{fig:QCDnLX} for small velocities the 
energy loss is larger at $\theta_v=\pi/2$ while for large velocities it is 
larger at $\theta_v=0$ similar to what we found at finite $\xi$. From 
Fig.~\ref{fig:QCDLX} we see that the energy loss of a bottom quark with 
$\theta_v=\pi/2$ is less than the isotropic result while the energy loss of a 
bottom quark with $\theta_v=0$ is larger than the isotropic result for 
$p\,\raisebox{-1mm}{$\stackrel{>}{\sim}$}\,10$~GeV; however, the results are not 
dramatically different than those obtained at $\xi=10$. However, the velocity 
where the energy loss becomes negative, $v\sim0.65$, is well within the range of 
applicability of the heavy quark approximation for a bottom quark ($v\geq0.39$). 
We are therefore led to the conclusion that for very strong anisotropies heavy 
quarks with small velocities will experience an energy gain rather than an 
energy loss.  Additionally we see that the energy loss is more negative at 
$\theta_v=0$ than any other angle of propagation.  This means particles moving 
along the anisotropy direction experience the most energy gain. This is a rather 
surprising result but its origin can be traced to the presence of modes on the 
unphysical sheet as discussed in Sec.~\ref{sec:umodes}.  These modes therefore 
represent a qualitative change to the physics in highly anisotropic plasmas.

\section{Conclusions}
\label{Conc}

In this paper we have calculated the complete leading-order collisional energy 
loss of a heavy quark propagating through an anisotropic quark-gluon plasma. The 
results were presented as a function of the heavy quark velocity and propagation 
angle.  We presented a discussion concerning the kinematical range over which 
the results obtained here apply when the quark mass is large but not infinite 
and we then explicitly applied the results to the energy loss of a bottom quark 
in an anisotropic setting. It was shown that for anisotropic systems the 
collisional energy loss has an angular dependence which increases as the 
coupling and/or anisotropy is increased as expected on intuitive grounds. 
Quantitatively, for $\alpha_s=0.3$ and a 20 GeV bottom quark we found that the 
deviations from the isotropic result were on the order of 10\% for $\xi=1$ and 
of the order of 20\% for $\xi \geq 10$.  When translated into the difference 
between longitudinal and transverse energy loss this results in a 10\% 
difference at $\xi=1$, a 30\% difference at $\xi=10$, and a 50\% difference at 
$\xi=\infty$.  

More importantly, however, we found that for small velocities the sign of the 
energy loss becomes \emph{negative} representing energy gain instead of loss 
whenever $\xi>0$. In the isotropic limit and assuming an infinitely heavy quark 
it can be shown that the energy loss is positive definite and our results 
confirm this expectation for isotropic systems in contrast to the earlier 
calculation of Braaten and Thoma.  In the anisotropic case, however, there is no 
guarantee that the energy loss will be a positive quantity.  We have shown here 
that the negative contributions to the energy loss come from singularities on 
the neighboring unphysical sheets which are not present in the isotropic case 
\cite{Romatschke:2004jh}.  For small anisotropies the velocities at which the 
energy loss becomes negative are lower than the thermal bound set by 
$v\sim\sqrt{3T/m}$.  For large anisotropies the velocities at which the energy 
loss becomes negative for a bottom quark are above the thermal lower bound and 
are therefore within the range of applicability of the infinite quark mass limit 
considered here.

The results we have presented for the heavy quark energy loss should be compared 
with the expected momentum-space anisotropy generated during the early stages of 
RHIC and LHC heavy-ion collisions.  At the earliest times that a particle 
description is appropriate the relevant scales are $p_T \sim Q_s$ where $Q_s$ is 
of order 1 GeV at RHIC and 2-3 GeV at LHC energies and $p_z \sim 1/\tau$ where 
$\tau$ is the time at which the hard gluon occupation number drops below 1. In 
order to estimate $\tau$ we follow the logic used in the bottom-up 
thermalization scenario \cite{BMSS:2001} giving parametrically $Q_s \tau \sim 
\alpha_s^{-3/2}$.  We therefore estimate that parametrically $p_z \sim 
\alpha_s^{3/2} Q_s$ and as a result $\xi \sim \alpha_s^{-3/2}$.  We therefore 
see that in the weak-coupling limit large anisotropies can be generated if the 
bottom-up scenario is true. Using a realistic coupling of $\alpha_s \sim 0.3$ 
gives $\xi \sim 6$ but this estimate could change dramatically depending on the 
specific coefficients omitted in our  parametric estimates. Additionally the 
time estimate $Q_s \tau \sim \alpha_s^{-3/2}$ is based on an implicit assumption 
of isotropy and collisional broadening and must be revisited in the context of 
an anisotropic quark-gluon plasma. Also in terms of producing phenomenologically 
relevant predictions it should be mentioned that since there are unstable modes 
present in the system, whatever value of $\xi$ one starts with, it is expected 
to go to zero very rapidly. In order to predict the total effect on observables 
we would need to fold together the results obtained here with the time evolution 
of $\xi$ which is not currently available.

It should be mentioned that thermal effects are only expected to make the energy 
loss \emph{more negative} at small velocities; however, it would be nice to have 
an explicit calculation of the anisotropic energy loss in the limit $v 
\rightarrow 0$ with the infinite-mass assumption relaxed. Additionally, a 
refinement of the calculation involving different momentum-space anisotropies 
for the quark and gluon distribution functions was suggested. Another area for 
improvement is that this work ignores next-to-leading order radiative energy 
loss.  In isotropic systems it has been shown that the radiative energy loss of 
a heavy quark is larger than the collisional energy loss and thus cannot be 
ignored. As illustrated by this work, however, a calculation of the radiative 
energy loss for an anisotropic system would be considerably more involved than 
the equivalent calculation in an isotropic one.  Nevertheless, it seems 
necessary in order to truly understand heavy quark energy loss in an anisotropic 
plasma.

\section*{Acknowledgments}
M.S. and P.R. would like to thank A.~Rebhan for discussions.  
M.S. was supported by the Austrian Science Fund Project No. M790.

\bibliography{TFTSlac.bib}
\bibliographystyle{utphys}

\end{document}